\documentclass[prd,  twocolumn, 10pt]{revtex4-1}

\usepackage{sirimacro}

\usepackage{amsmath}
\usepackage{amsfonts}
\usepackage{amssymb}
\usepackage{graphicx}
\usepackage{epsfig}
\usepackage{mathrsfs}
\usepackage{bm}
\usepackage{microtype}
\usepackage{hyperref}
\usepackage{mathrsfs}

\newcommand{\LL}{^\ell\lambda_H}
\newcommand\HH{\mathscr{H}}
\renewcommand\eps\epsilon
\def\tdN{\tilde{N}}

\begin{document}

\title{Inflation model building with an accurate measure of e-folding}

\author{Sirichai Chongchitnan}

\date{\today}
\affiliation{E. A. Milne Centre for Astrophysics, University of Hull,  Cottingham Rd., Hull, HU6 7RX, United Kingdom.}

\email{s.chongchitnan@hull.ac.uk}

\begin{abstract}
It has become standard practice to take the logarithmic growth of the scale factor as a measure of the amount of inflation, despite the well-known fact that this is only an approximation for the true amount of inflation required to solve the horizon and flatness problems. The aim of this work is to show how this approximation can be completely avoided using an alternative framework for inflation model building. We show that using the inverse Hubble radius, $\HH=aH$, as the key dynamical parameter, the correct number of e-folding arises naturally as a measure of inflation.  As an application, we present an interesting model in which the entire inflationary dynamics can be solved analytically and exactly, and, in special cases, reduces to the familiar class of power-law models.
\end{abstract}

\maketitle
\section{Introduction}

The horizon problem was a long-standing conundrum identified during the early development of modern cosmology \cite{misner}. The problem was the observation of large-scale homogeneity - even amongst regions that would not have been in causal contact with one another under the assumption of Friedmann-Robertson-Walker cosmology.

A mechanism which could solve the horizon problem (and other cosmological problems), is inflation \cite{starobinsky2,guth,linde}. Inflation stipulates that  the early Universe expanded in a brief de-Sitter phase, leading to the loss of causal contact between regions that were previously causally connected. Indeed, inflation is defined as the period during which the comoving Hubble radius of the Universe \ii{shrinks}, \ie
\ba\diff{}{t}\bkt{1\over aH}<0\lab{end}\ea

The factor by which the Hubble radius shrinks is parametrized as $e^{\td N}$ where ${\td N}$ is the  number of e-folds (the reason for the tilde will soon be apparent). Thus, by definition,
\ba\td{N}(t)\equiv \ln{a(t\sub{end})H(t\sub{end})\over a(t)H(t)}.\lab{two}\ea

However, in the simplest `slow roll' models of inflation, the Hubble parameter, $H$, is essentially constant compared to  the scale factor, $a$, which increases exponentially. For these models, it is adequate to quantify the `amount' of inflation using a simpler e-fold number, $N$, which measures the growth of the scale factor alone, \ie
\ba N(t)\equiv \ln{a(t\sub{end})\over a(t)}= \td{N}+\left|\ln{H(t)\over H(t\sub{end})}\right|.\lab{only}\ea



To distinguish the two definition, we will refer to $\td N$ as the number of \ii{physical} e-folds. Note that $N=\tdN=0$ at the end of inflation, and at any given $t<t\sub{end}$, we have $\td{N}<N$.

Whilst it has become industry standard to use $N$ to quantify the amount of inflation, $N$ does not directly quantify how much inflation alters the causally connected volume of the Universe, as the causal volume is of course defined by the Hubble radius and not the scale factor. Liddle, Parsons and Barrow \cite{liddle+} were the first to emphasise that the physical e-folds $\td N$ is a more accurate measure of inflation.  A number of authors have subsequently echoed this sentiment \cite{dodelson,LL,delCampo}.

In this work, we address the following questions: 
\bee
\item For what kind of inflation models is $N$ not an accurate measure of inflation, and how can this inaccuracy be quantified?
\item If $\td N$ were used to quantify inflation instead of $N$, how would this affect the predictions for inflationary observables (like the tensor-to-scalar ratio)? Is this change significant for future experiments? 
\item  (the primary aim of this paper) How can we construct inflation models in which $\td N$ arises naturally as the parameter which quantifies the amount of inflation?
\eee

Section \ref{patho} addresses points (1) and (2). Section \ref{secHH} discusses point (3). Throughout this work,  we measure $\phi$ in units of the reduced Planck mass, \iee we set $m\sub{pl}/\sqrt{8\pi}\equiv1$.

\section{Numerical investigation}\lab{patho}

We are searching for inflation models  in which $H$ varies sufficiently for $N$ and $\td N$ to differ significantly. In these models, the conventional slow-roll wisdom does not apply (this will become clear shortly). To investigate generic properties of such models, we will construct models stochastically by  employing the Hamilton-Jacobi formulation which we now briefly describe. See \cite{lidseybig,delCampo} for reviews of this formalism.

\subsection{The Hamilton-Jacobi formalism}

In the Hamilton-Jacobi (HJ) formalism, it is the Hubble parameter, $H$, rather than the inflaton potential, $V$, which plays the central role in determining  the dynamics of inflation. The two parameters are related by the Hamilton-Jacobi equation and the Friedmann equation, given respectively by
\ba
\left(H'(\phi)\right)^2-{3\over2}H^2(\phi)&=-{1\over2}V(\phi)\label{ham},\\
 H^2&= {1\over3}\bkts{{\dot\phi^2\over2}+V(\phi)},\lab{fried}\ea
where $\phi$ is the inflaton field value. It is clear from these equations that  the approximation $H(\phi)\approx$  constant is equivalent to the slow-roll limit $\dot\phi^2\ll V(\phi)$. The HJ formalism has the advantage that the dynamics of the inflaton can be analysed exactly without slow-roll approximation.

\subsection{Dynamics of stochastically generated models}

To gain a generic quantitative understanding of how $\td N$ affects the dynamics of inflation, we will generate models of inflation stochastically. 

We follow the same setup as the `flow-equation' approach previously studied in \cite{kinney,me1,hoffman,ramirez}, amongst others (although the flow equations themselves - a large set of coupled ODEs, will not be required). In this approach, $H(\phi)$ is a polynomial of an arbitrarily high  degree, with coefficients drawn from a uniform set of priors. In other words, we write
\ba H(\phi)&= H(0)\bkts{1+C_1^0\phi+{C_2^0\over 2!}\phi^2+{C_3^0\over 3!}\phi^3\ldots},\notag\\
C_n^0 &\equiv {H^{(n)}(0)\over H(0)}, \lab{tayH}\ea
and pick $C_n^0$ randomly from some priors, with the terminating condition $C_n^0=0$ for $n>N$. Once $H(\phi)$ is chosen, we evolve $H$ via the HJ equation and determine whether the model is a viable candidate for inflation. 

We will use 0 superscript to denote values at $\phi=0$. We can also define a more general form of $C_n$ as the scaled $n\super{th}$ derivative of $H$ with respect to $\phi$:
\bas C_n &\equiv {H^{(n)}(\phi)\over H(\phi)}. \eas

The $C_n$ are related to the so-called Hubble `slow-roll' (HSR)  parameters (even though no slow roll is assumed). The HSR parameters are:
\ba \epsilon &\equiv 2\left({H'\over H}\right)^2,\qquad \eta \equiv2{H''\over H}\ff,\nn\\ 
\sigma &\equiv 2\eta -4\epsilon\ff,\qquad \ff\ff\xi\equiv4{ H^\prime H^{\prime\prime\prime}\over
H^2}\ff,\nn\\
^\ell\lambda_H\ &\equiv 2^\ell{(H')^{\ell-1}\over H^\ell}{d^{\ell+1}H\over
d^{\ell+1}\phi}\ff\label{flowparam}. \ea
The relations between the two sets of parameters at $\phi=0$ are
\ba C_1^0 &= -\sqrt{\eps_0/2}\notag\\
C_2^0 & = \eps_0+\sigma_0/4\notag\\
C_n^0 &= {(-1)^n {(^{n-1}\lambda_{H,0})}\over 2(2\eps_0)^{(n-2)/2}},\quad 3\leq n\leq n\sub{max}.\lab{genf}
\ea
These results were partially obtained in \cite{ramirez} (our formula \re{genf} generalises their results and corrects a sign error in that paper). 
The priors we will use to generate inflation models are 
\ba \epsilon_0&\in[0,0.8]\ff,\nn \\\sigma_0&\in[-0.5,0.5]\ff,\nn \\{\xi}_0&\in[-0.05,0.05]\ff, \label{window}\\ \LL|_0&\in[-0.025\times5^{-\ell+3},0.025\times5^{-\ell+3}]\ff, \quad 4\leq\ell\leq n\sub{max}-1\nn \ea



Once $H(\phi)$ is chosen, it remains to determine the range of $\phi$ in which inflation takes place. In particular, we will need to know the value $\phi\sub{end}$ at which inflation ends, and $\phi=\phi_*$ where CMB-scale perturbations were generated.

By definition, Inflation occurs as long as \re{end} is satisfied. This is equivalent to the condition 
$$\eps<1.$$
In a randomly generated model,  the root of the equation $\eps(\phi)=1$ may not exist.  These are models that will require an additional mechanism to end inflation (such as hybrid inflation with a secondary field). We disregard these models in this work.  


To calculate inflationary observables, we need to determine the field value, $\phi_*$, corresponding to the moment when CMB-scale $k$ mode first exited the Hubble radius. In this section, we will take this so-called `horizon-exit' to be at  60 e-folds before the end of inflation. 

If the conventional e-fold $N$ is used, then we determine the root, $\phi_*$, from the integral equation
\ba \int_{\phi_*}^{\phi\sub{end}} {\D \phi\over \sqrt{2\eps}}=60. \ea
This follows from differentiating the definition \footnote{We will assume that $\phi(t)$ increases monotonically during inflation, and since $N$ and $t$ run in opposite directions, \re{solve-n} implies that $H^\pr(\phi)<0$ during inflation.} of $N$:
\ba \diff{N}{\phi}={H\over2H^\pr}.\lab{solve-n}\ea

If the physical e-fold $\td N$ is used, then we need to solve a more complicated equation:
\ba \int_{\phi_*}^{\phi\sub{end}} {\D \phi\over \sqrt{2\eps}}-\ln{H(\phi_*)\over H(\phi\sub{end})}=60. \ea

Once $\phi_*$ is obtained, we can then calculate the inflationary observables.

\subsection{The $n_S$-$r$ plane}

Two inflationary observables that are most strongly constrained by CMB temperature anisotropies and polarization are the tensor-to-scalar ratio, $r$, and the scalar spectral index, $n_S$. To calculate these observables for our stochastic models, we use the next-to-leading order expressions for derived in \cite{lidseybig} 
\ba r &\simeq 16\epsilon[1-C(\sigma+2\epsilon)]\ff, \label{r2}\\ 
n_S&\simeq1+\sigma-(5-3C)\epsilon^2-{1\over4}(3-5C)\sigma\epsilon  +{1\over2}(3-C)\xi, \label{ns2}
\ea 
where  $C=4(\ln2+\gamma)-5\simeq0.0814514$ (with $\gamma$ the Euler-Mascheroni constant). 

\begin{figure*}[htbp]    \centering
   \includegraphics[width=\textwidth]{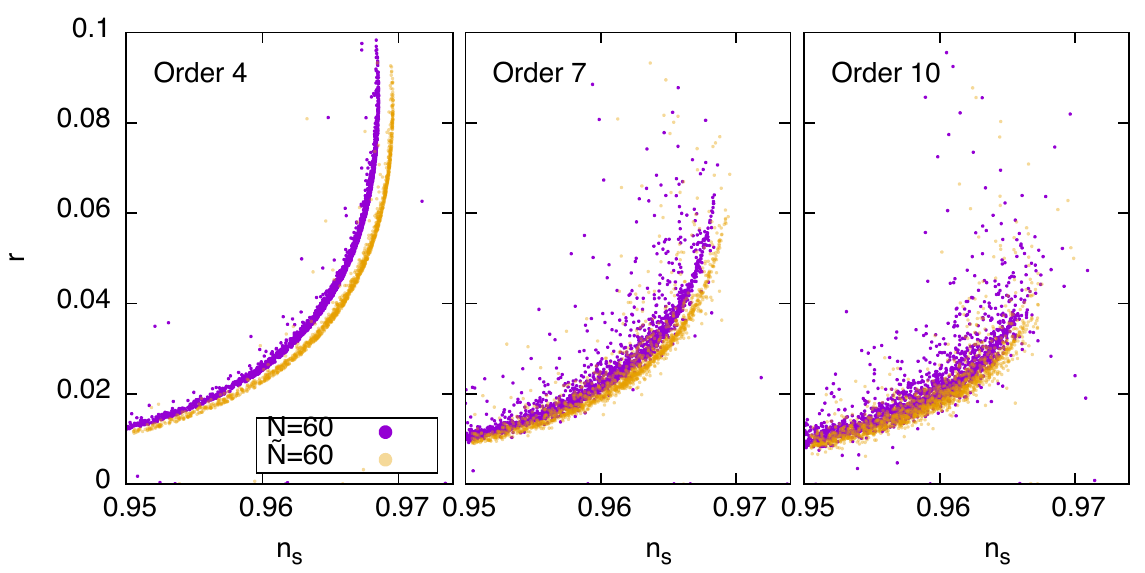} 
   \caption{The $n_S$-$r$ plane where $H(\phi)$ is a randomly generated polynomial of varying order (4,7 and 10 from left to right). The dark/purple swathes are generated assuming the approximation $N=60$, whereas the light/orange swathes assume the correct definition of e-folding: $\td N=60$.}
   \label{fig_trip}
\end{figure*}

The stochastic search produces a structure in the $n_S$-$r$ plane as shown in Fig. \ref{fig_trip}. Each of the 3 panels shows 2000 models where $H(\phi)$ is assumed to be a polynomial of degree 4, 7 or 10. Purple/dark points are the predictions for those models for which $N=60$, whereas the orange/light points are those for the correct requirement  $\td N=60$. We only consider models which satisfy the bounds
\ba r\leq0.1 \quad -0.95\leq n_S\leq 0.98\ea
corresponding roughly to those from Planck \cite{planck15}.

The swathes for the physical e-fold generally correspond to lower values of $r$. One way to understand this is as follows. From Eq. \ref{only}, requiring that $\td N=60$ means that CMB-scale perturbations were generated at $N>60$. Our previous phase-space analysis showed that as $N\to\infty$, we have the limit 
\ba\lim_{N\to\infty}\eps=0.\ea 
We refer the reader to \cite{me1} for detailed analysis of the dynamics in this plane. This means that $r$ also approaches $0$ as $N\to\infty$. This explains why the $\td N=60$ swathes are displaced towards lower $r$.

In the same work, we also showed that as $N$ increases, $\sigma$ generally approaches 0 from below, meaning that $n_S<1$, but approaches scale invariance as $N\to\infty$. Therefore, $n_S$ at $\td N=60$ will generically be larger than that with $N=60$. This explains why the $\td N=60$ swathes are also shifted to the right.

It is possible to construct $H(\phi)$ which do not conform to the above explanation, although they require a conspiratorial combination of the coefficients $C_n$. We will not analyse these cases here, but simply note that they do not arise easily out of such a stochastic framework.

We note that increasing the order of the polynomial generally dilutes the density of the swathes (for a fixed total number of models). Increasing the order to beyond 10 produces very similar swathes to those for order 10.

\begin{figure*}[htbp]    \centering
   \includegraphics[width=\textwidth]{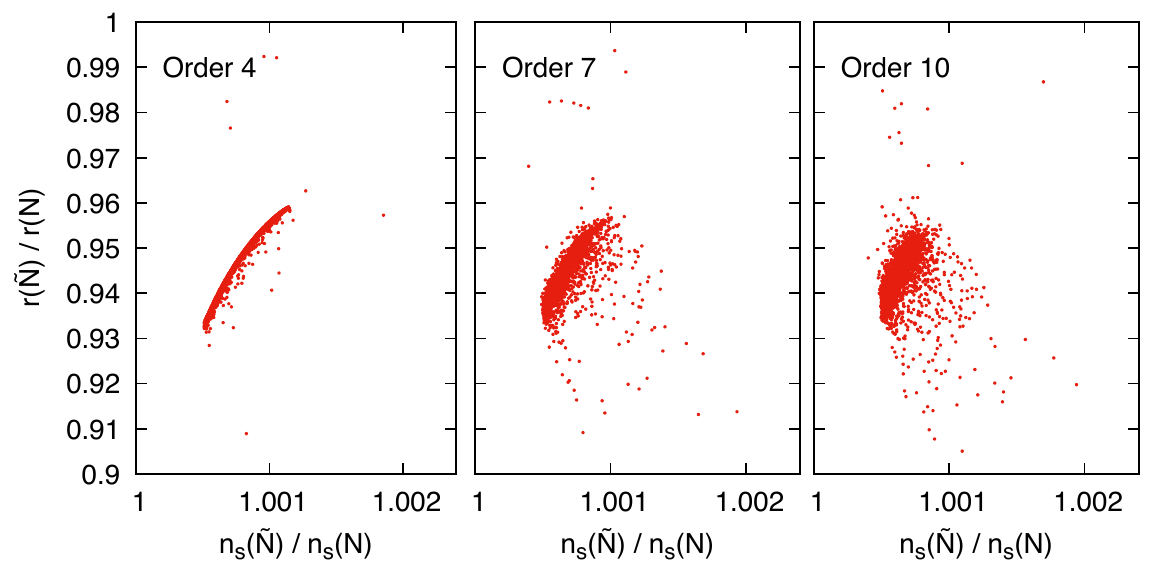} 
   \caption{The ratio between the $(n_S, r)$ values of the two swathes in each panel in Fig. \ref{fig_trip}, \ie{} $r(\tilde N=60)/r(N=60)$ plotted against $n_S(\tilde N=60)/n_S(N=60)$. }
   \label{fig-rat}
\end{figure*}

Next, we quantify the observed downwards and rightwards shifts in terms of the ratio between the observables for the two e-fold definitions. Figure \ref{fig-rat} shows this result. When using $\td{N}=60$ instead of $N=60$, we find that $r$ is decreased by $\lesssim10\%$ and $n_S$ is increased by $\sim1\%$. The dynamical profiles of the outliers in this figures will now be investigated.

We point out that these corrections  can be calculated for any form $H(\phi)$. We have only dealt with polynomials in this work, and it is possible to construct examples with more extreme corrections using other forms of $H(\phi)$ such as the Pad\'e series \cite{ramirez,liddle+, coone}. We will investigate this in future work.

\subsection{Profile of pathological models}

We now take a closer look at the dynamics of inflation models that are, and are not, sensitive to the choice of e-fold definition. 

\begin{figure*} 
   \centering
   \includegraphics[width=0.49\textwidth]{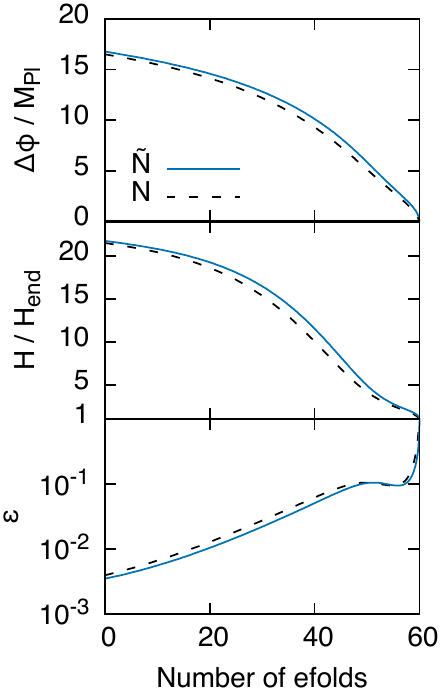}  \includegraphics[width=0.5\textwidth]{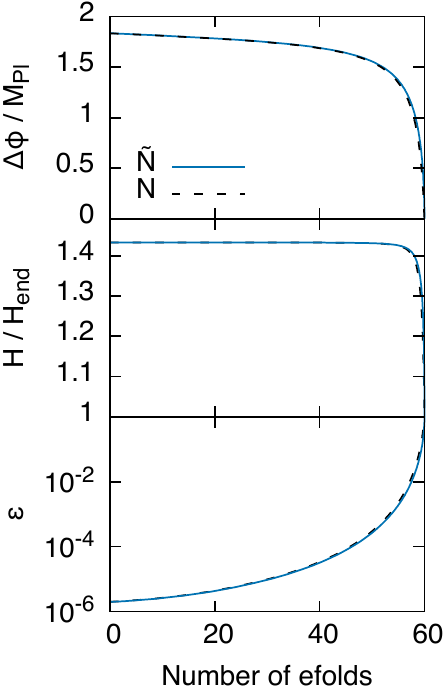} 
   \caption{The dynamical evolution of two inflation models (left and right panels). For each mode, we plot the evolution of $\Delta\phi$ (field excursion), Hubble parameter $H$ and the Hubble `slow-roll' parameter, $\eps$, as a function of the total number of e-folds measured from horizon exit (60 marks the end of inflation). Solid/blue lines use $\td N$ as the measure of e-folds, whilst dashed/black lines use the approximation $N$.  Although the dynamics of the model on the right is essentially unchanged whether $N$ or $\td N$ is used, the model on the left is inaccurately measured by $N$. In this particular case, $r(\tdN=60)$ is 12\% smaller than $r(N=60)$.}
   \label{long_fig}
\end{figure*}

In particular, we consider the evolution of 1) the excursion of the inflaton field, $\Delta\phi$, 2) the Hubble parameter, $H$, and 3) the HSR parameter, $\eps$. These quantities are plotted in Fig. \ref{long_fig} for two particular models that illustrate typical behaviours of models that are sensitive (left panels) or insensitive (right) to the e-fold definition. The curves are plotted as a function of the total number of e-folding $\Delta N$ (or $\Delta\tdN$) measured with respect to horizon-exit (so that 60 marks the end of inflation). Solid (dashed) line shows the evolution with respect to $\td N$ ($N$). The left vertical panel shows the dynamics for a model in which $r(\td N)/r(N)$ is large (we will refer to this kind of model as `pathological'), whereas the other panel corresponds to a typical slow-roll model in which the observables are insensitive to whether  $N$ or $\td N$ is used (the two types of lines are almost indistinguishable in this panel).

We observe the following behaviour for the pathological models. 
\bit
\item The field excursion is large ($\Delta\phi\sim10$). For slow-roll models, $\Delta\phi\sim1$.
\item The Hubble parameter varies over a wide range of values, whereas $H$ is approximately a constant during most of the inflationary period.
\item The HSR $\eps$ is, on average, many times larger than the slow-roll case.
\eit
Note that these are generic observations in the class of polynomial $H(\phi)$, and not restricted to these two specific models.

\subsection{Caveats for the flow equations}

In the majority of previous work surveying inflation models using the HJ approach, $H(\phi)$ is not solved directly. Instead, the inflaton dynamics with respect to $\phi$ is eschewed in favour of that with respect to $N$. This method was first formalised by Kinney \cite{kinney}, who showed that the inflationary dynamics is determined by solving a system of coupled ODEs for the HSR variables:
\ba \diff{\epsilon}{ N}&= \epsilon(\sigma+2\epsilon)\ff,\nn\\
\diff{\sigma}{N} &= -\epsilon(5\sigma+12\epsilon)+2(\ff^2\lambda_H)\ff,\\
\diff{}{N}\ff^\ell\lambda_H &= \Big[{\ell-1 \over 2}\sigma +(\ell-2)\epsilon\Big]\ff^\ell\lambda_H+ \ff^{\ell+1}\lambda_H\ff  \ff(\ell\geq2).\nn
\ea
These are the so-called `inflationary flow equations'. 

Dodelson and Hui \cite{dodelson} suggested that a similar set of flow equations with respect to $\td N$ could be obtained, since the differentiation in $N$ and $\td N$ are simply related by 
\ba \diff{}{\bar N}={1\over  1-\eps}\,\diff{}{N},\lab{possible}\ea
although they did not implement this explicitly.

But there is an obvious problem with this suggestion: the fact that as $\eps\to1$ towards the end of inflation, the ODEs blow up. Thus, the flow equations are unsuitable for the study of dynamics with respect to the physical e-folds (see \cite{vennin} for other problems with the flow equations).

In our stochastic implementation (based on the observation of Liddle \cite{Liddle}), there is no  integration of coupled ODEs. Instead, we have to tackle a couple of root-finding problems (\ie{} solving for $\phi\sub{end}$ and $\phi_*$). This approach reveals inflationary dynamics with respect to $\td N$ that cannot be obtained using the flow equations.



\section{A new approach to inflation model building}\lab{secHH}

We now describe a new approach to inflation model building in which $\td{N}$ arises as the natural measure of the amount of inflation. By `natural' we mean that the dynamics with respect to $\tdN$ are obtained much more easily than that in $N$. We will illustrate our method with an example of an observationally-viable class of models whose dynamics can be solved exactly by hand.


\subsection{$\HH(\phi)$}



We propose taking the comoving Hubble parameter $\mathscr{H}=aH$ (the inverse Hubble radius) as the fundamental parameter in place of $H(\phi)$. In analogy to Eq. \ref{tayH}, we begin with the Taylor series of $\mathscr{H}$ about $\phi=0$:
\ba \HH(\phi)&= \HH(0)\bkts{1+E_1^0\phi+{E_2^0\over 2!}\phi^2+{E_3^0\over 3!}\phi^3\ldots},\\
E_n^0 &\equiv {\HH^{(n)}(0)\over \HH(0)}. \ea
As before, we use the superscript $0$ to differentiate these constant coefficients from the field-dependent quantities 
\ba E_n (\phi) &\equiv {\HH^{(n)}(\phi)\over \HH(\phi)} .\ea
By definition, inflation occurs as long as 
$${\D\over \D t}\HH>0.$$ 
Assuming $\phi(t)$ increases monotonically, the inflation  condition becomes 
\ba\HH^\pr(\phi)>0\iff E_1>0,\ea which are the equivalent to the exact condition $\eps<1$ in the HJ formalism. The difference is that whilst $H(\phi)$ is a decreasing function during inflation, $\HH(\phi)$ is an increasing function.

In fact, $E_1(\phi)$ has a straightforward interpretation as the (negative of the) infinitesimal e-folding that occurs as $\phi$ evolves. This can be seen from differentiating the definition of $\td N$.
\ba\diff{\tilde N}{\phi}=-{\mathscr{H}^\pr\over\mathscr{H}}=-E_1.\lab{solve-ana}\ea
It is instructive to compare the above equation with Eq. \ref{solve-n}, for which there is no general analytic solution, whereas the solution to \re{solve-ana} is simply Eq. \ref{two}, which can now be written as
\ba \td N(\phi)=\ln\bkt{\HH(\phi\sub{end})\over\HH(\phi))},\lab{hnd}\ea
where $\phi\sub{end}$ is the solution of $\HH^\pr(\phi)=0$ (in other words, inflation ends at a turning point of $\HH(\phi)$). Setting $\td N=60$, we can then solve for $\phi_*$,

Finally, to calculate the observables, we obtain the following expressions for the HSR parameters  in terms of $E_n^*$ after some algebra.
\ba \eps  &= {2\over\bkt{E_1^*+\sqrt{(E_1^*)^2+2}}^2},\\
\eta &= {\eps(2E_2^* + 3) -1\over 1+\eps},\\
\xi &= {\eps\over (1+\eps)^3}\bigg(3\eps^3-2\sqrt2\eps^{5/2}E^*_3-2E^*_2\eps^2+8\eps (E_2^*)^2-3\eps^2 \ldots\notag\\
&-4\sqrt2E^*_3\eps^{3/2}+28\eps E^*_2+17\eps-2\sqrt{2}E^*_3\sqrt{\eps}-2E^*_2 -9 \bigg).
\ea
These can then be substituted into the formulae for $r$ and $n_S$ \re{r2}--\re{ns2} as before. 



\subsection{A simple example}
Let us illustrate the $\HH(\phi)$ formalism with the Gaussian model
\ba \HH(\phi)= \HH(\phi\sub{end})\,e^{-\alpha^2\phi^2},\quad \alpha>0.\ea 

Inflation occurs on the branch where $\HH$ is increasing, \iee $\phi<0 \,$\footnote{The negative $\phi$ value is not problematic due to the even symmetry of $\HH$ and the $t\to-t$ transformation.} and ends at $\phi\sub{end}=0$ (where $\HH^\pr=0$). 

\begin{figure}[htbp]    \centering
   \includegraphics[width=\columnwidth]{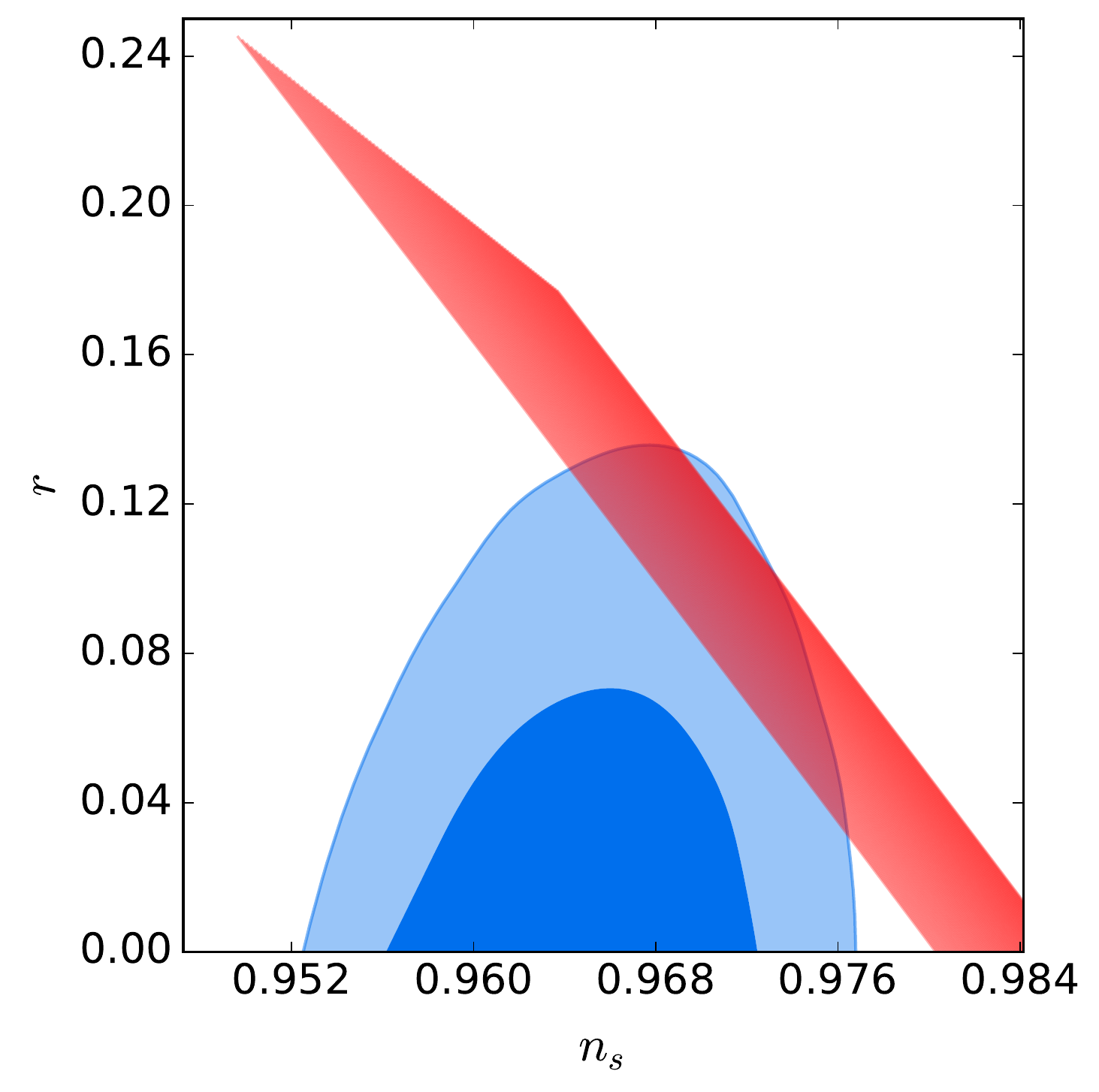} 
   \caption{The $n_S$-$r$ prediction for the Gaussian model $\HH \propto \exp(-\alpha^2\phi^2)$ shown as the sharp diagonal, with parameter $\log_{10}\alpha\in[-0.4,5]$ (from top to bottom of the diagonal), assuming CMB-scale perturbations were generated at physical e-folds $\td{N}_*\in[50,70]$ (from left to right). We also show the  $2\sigma$ constraints in this plane from Planck \cite{planck15}, using both the temperature and low-$\ell$ polarization data. In fact, the diagonal coincides with the prediction for power-law potentials $V\sim\phi^n$, as discussed in \S\ref{surp}. }
   \label{figexpo}
\end{figure}

Using Eq. \ref{solve-ana}, and under a more general assumption that CMB perturbations were generated at $\td N=\td{N}_*$, we find the following:
\bas 
\phi_*&=-\alpha^{-1}\sqrt{\tdN_*},\qquad\qquad E_1^*=2\alpha\sqrt{\tdN_*},\\
E_2^*&=2\alpha^2(2\tdN_*-1),\ff\ff\quad E_3^*=4\alpha^3\sqrt{\tdN_*}(2\tdN_*-3).
\eas
These translate to the observables (to leading order):
\bas 
r\approx {2\over \alpha^2\tdN_* },\qquad n_S\approx 1-{r\over8}(1+4\alpha^2).
\eas
Figure \ref{figexpo} shows the predictions of this model in the $n_S$-$r$ plane for $\log_{10}\alpha\in[-0.4,5]$ and $\td{N}_*\in[50,70]$, superimposed on the Planck $2\sigma$ constraints from both the temperature and the low $\ell$ polarization data \cite{planck15}. With $\td{N}_*=60$, we find that the Planck constraints imply a lower bound $\alpha\gtrsim0.5.$

\subsection{Comparing $N$ and $\tdN$}
Using Eq. \re{possible}, it is possible to perform a direct comparison of $N$ and $\td N$ given a $\HH(\phi)$ model. In the case of the Gaussian model, \re{possible} can be integrated exactly to give:
\ba N= {\tdN\over2}\bkt{1+\sqrt{1 +{1\over 2\alpha^2\tdN}}}+{1\over 4\alpha^2}\sinh^{-1}\bkt{\alpha\sqrt{2\tdN}}.\ea

It is easy to show that  $N\to\tdN$ as $\alpha\to\infty$, but for small $\alpha$,  the number of physical e-folds could be rather different from the naive slow-roll expectation. 

\begin{figure}[htbp]    \centering
   \includegraphics[width=\columnwidth]{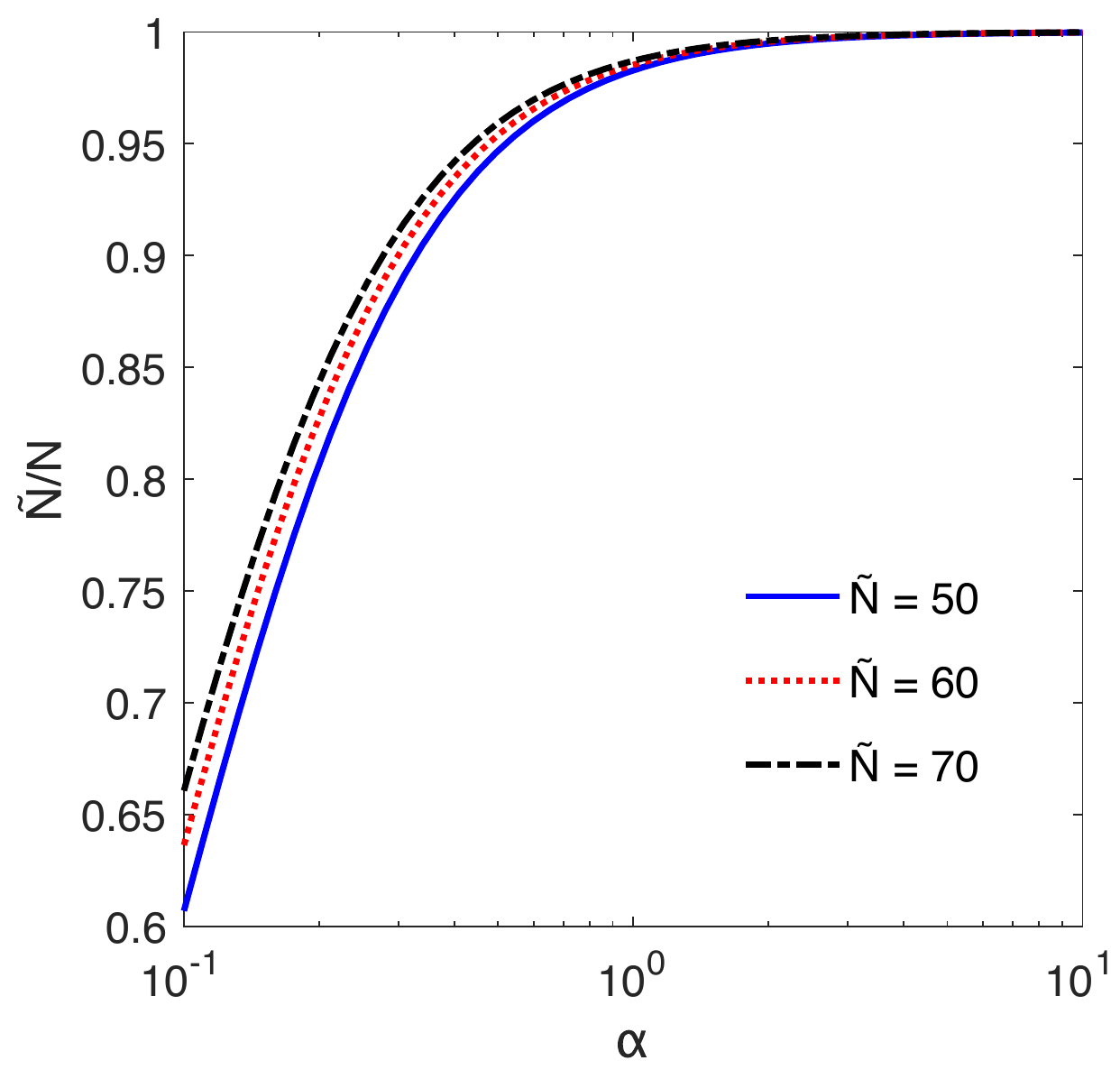} 
   \caption{The discrepancy between the physical and slow-roll e-folds for the Gaussian model plotted as a function of the parameter $\alpha$, assuming $\tdN=50-70$.}
   \label{figNN}
\end{figure}

Fig. \ref{figNN} shows this discrepancy (expressed as $\tdN/N$) as a function of $\alpha$ for $\tdN=50-70$. In this range of $\alpha$, the physical e-fold $\tdN$ could be as small as $60\%$ of the slow-roll e-fold. Thankfully, for values of $\alpha$ that are consistent with the Planck $2\sigma$ constraints, the maximum discrepancy reduces to about $5\%$. The latter translates to a correction in $r$ of a few percent.

\subsection{Connection to the potential}\lab{connex}

How does a given $\HH(\phi)$ correspond to an inflaton potential $V(\phi)$? We provide a flow chart for get from $\HH(\phi)$ to $V(\phi)$ below.
\ba \HH(\phi)  \Longrightarrow E_1=\HH^\pr/\HH  \Longrightarrow\ff\sqrt{\eps} = &{\sqrt{E_1^2+2}-E_1 \over \sqrt2}\notag\\
&\big\Downarrow\\
 V(\phi)=H^2(3-\eps) \ff \Longleftarrow \ff H(\phi)=H\sub{end}&\exp\bkt{-\int_{\phi\sub{end}}^\phi\sqrt{\eps/2}\D\phi}\notag 
\ea
This procedure will produce $V(\phi)$ \ii{exactly}. In the case when slow-roll holds to a good approximation, $V(\phi)$ can be obtained directly from $\eps$ by the integration:
\bas V\approx\exp\bkt{\int\sqrt{2\eps}\D\phi}\eas

For example, take $\HH=$ constant. Following the flow chart, we have the correspondence to the quintessence-type exponential potential:
\bas \HH(\phi) \sim \text{constant} \iff V(\phi)\sim e^{-\sqrt2\phi}.\eas

\subsection{A surprising correspondence}\lab{surp}
For the Gaussian $\HH(\phi)$, it turns out that we have a surprising correspondence to the power-law potentials $\phi^n$:
\ba \HH(\phi) \sim e^{-\alpha^2\phi^2} \iff V(\phi)\sim \phi^n  \ff\text{ where } n={1\over2\alpha^2}\lab{magic}\ea
An easy way to to see this correspondence is to consider the expression for $\eps(\tdN)$
\ba \eps(\td N)&={1\over1+4\alpha^2\tdN+2\alpha\sqrt{2\td N(2\alpha^2\tdN+1)}}\notag\\
&\approx  {1\over 1+8\alpha^2\tdN}. \lab{ohwow}\ea
This approximation holds as long as $2\alpha^2\tdN\gg1$, which is valid in the ranges of $\alpha$ and $N$ that are consistent with Planck's constraints. Comparing \re{ohwow} with the slow-roll expression for $V(\phi)\propto \phi^n$ (see for example \cite{planck15}), 
\ba \eps_V(N)\approx{1\over1+4N/n }\ea
and assuming the slow-roll limit $\tdN\approx N$, we find the correspondence
\ba n={1\over 2\alpha^2}.\ea
Indeed the red diagonal in Fig. \ref{figexpo} matches the predictions for $\phi^n$ with $N\in[50,70]$ almost exactly. 

In future work, we will use this and other correspondences between $V(\phi)$ and $\HH(\phi)$ to help elucidate the link between $V(\phi)$ and $H(\phi)$ via the HJ equation, which is notoriously difficult to solve analytically.




\section{Summary and discussion}

Statements such as ``$N=60$" in fact underestimate the actual  number of physical e-folds needed to solve the horizon problem. The severity of the discrepancy increases with the departure from the  slow-roll condition, as is well known. In this work, we quantified this discrepancy for a large class of models, and showed how to avoid such an approximation using an alternative modeling of inflation.

Our main results can be summarised as follows. 
\bit
\item For the class of inflation models parametrized as polynomial $H(\phi)$ (previously studied in the context of the inflationary flow equations), using the `wrong' definition of e-fold could lead to the over-prediction of the tensor-to-scalar ratio $r$ by about  $10\%$.
\item We propose modelling inflation using $\HH=aH$ (the inverse Hubble radius). In this approach, the physical e-fold $\tdN$ arises naturally as a  measure of the amount of inflation. 

\item In the Hamilton-Jacobi formalism, Eq. $\re{solve-n}$ links the e-fold dynamics to the field dynamics, but the equation can rarely be solved exactly, In contrast, our $\HH(\phi)$ approach easily provides such a link via Eq. \ref{hnd} (where the RHS is known). There are no difficult ODEs to solve.

\item As an application, we showed that the Gaussian function, $\HH(\phi)=e^{-\alpha^2\phi^2}$, reduces to the power-law potential $V(\phi)\sim\phi^n$ in the observationally-relevant parameter space. We showed that in this case the discrepancy between $\tdN$ and $N$ can be obtained analytically. Under Planck's $2\sigma$ constraints on $r$ and $n_S$, the discrepancy could be of order $5\%$.
\eit

Whilst the magnitude of corrections presented in this work may not seem significant, in our opinion it seems entirely unnecessary  to invoke the approximation  $N\approx \td N$ when analysing inflation models. It compromises not only the principle of  \ii{exactness} in the HJ approach, but is contrary to drive towards precision cosmology. The $\HH(\phi)$ approach completely avoids this problem.

The suite of upcoming ambitious experiments, including CMB polarization experiments and direct gravitational wave observatories, will be able to measure inflationary observables on very different physical scales, and therefore constrain models that go beyond the conventional (convenient?) slow-roll wisdom that $H$ is constant during inflation. We believe that the formalism presented here will be useful in this endeavour.






\bibliographystyle{apsrev4-1}
\bibliography{inflation}

\begin{thebibliography}{19}%
\makeatletter
\providecommand \@ifxundefined [1]{%
 \@ifx{#1\undefined}
}%
\providecommand \@ifnum [1]{%
 \ifnum #1\expandafter \@firstoftwo
 \else \expandafter \@secondoftwo
 \fi
}%
\providecommand \@ifx [1]{%
 \ifx #1\expandafter \@firstoftwo
 \else \expandafter \@secondoftwo
 \fi
}%
\providecommand \natexlab [1]{#1}%
\providecommand \enquote  [1]{``#1''}%
\providecommand \bibnamefont  [1]{#1}%
\providecommand \bibfnamefont [1]{#1}%
\providecommand \citenamefont [1]{#1}%
\providecommand \href@noop [0]{\@secondoftwo}%
\providecommand \href [0]{\begingroup \@sanitize@url \@href}%
\providecommand \@href[1]{\@@startlink{#1}\@@href}%
\providecommand \@@href[1]{\endgroup#1\@@endlink}%
\providecommand \@sanitize@url [0]{\catcode `\\12\catcode `\$12\catcode
  `\&12\catcode `\#12\catcode `\^12\catcode `\_12\catcode `\%12\relax}%
\providecommand \@@startlink[1]{}%
\providecommand \@@endlink[0]{}%
\providecommand \url  [0]{\begingroup\@sanitize@url \@url }%
\providecommand \@url [1]{\endgroup\@href {#1}{\urlprefix }}%
\providecommand \urlprefix  [0]{URL }%
\providecommand \Eprint [0]{\href }%
\providecommand \doibase [0]{http://dx.doi.org/}%
\providecommand \selectlanguage [0]{\@gobble}%
\providecommand \bibinfo  [0]{\@secondoftwo}%
\providecommand \bibfield  [0]{\@secondoftwo}%
\providecommand \translation [1]{[#1]}%
\providecommand \BibitemOpen [0]{}%
\providecommand \bibitemStop [0]{}%
\providecommand \bibitemNoStop [0]{.\EOS\space}%
\providecommand \EOS [0]{\spacefactor3000\relax}%
\providecommand \BibitemShut  [1]{\csname bibitem#1\endcsname}%
\let\auto@bib@innerbib\@empty
\bibitem [{\citenamefont {Misner}(1969)}]{misner}%
  \BibitemOpen
  \bibfield  {author} {\bibinfo {author} {\bibfnamefont {C.~W.}\ \bibnamefont
  {Misner}},\ }\href {\doibase 10.1103/PhysRevLett.22.1071} {\bibfield
  {journal} {\bibinfo  {journal} {Phys. Rev. Lett.}\ }\textbf {\bibinfo
  {volume} {22}},\ \bibinfo {pages} {1071} (\bibinfo {year}
  {1969})}\BibitemShut {NoStop}%
\bibitem [{\citenamefont {Starobinsky}(1982)}]{starobinsky2}%
  \BibitemOpen
  \bibfield  {author} {\bibinfo {author} {\bibfnamefont {A.~A.}\ \bibnamefont
  {Starobinsky}},\ }\href@noop {} {\bibfield  {journal} {\bibinfo  {journal}
  {Phys. Lett. B}\ }\textbf {\bibinfo {volume} {117}},\ \bibinfo {pages} {175}
  (\bibinfo {year} {1982})}\BibitemShut {NoStop}%
\bibitem [{\citenamefont {Guth}(1981)}]{guth}%
  \BibitemOpen
  \bibfield  {author} {\bibinfo {author} {\bibfnamefont {A.~H.}\ \bibnamefont
  {Guth}},\ }\href@noop {} {\bibfield  {journal} {\bibinfo  {journal} {Phys.
  Rev.}\ }\textbf {\bibinfo {volume} {D23}},\ \bibinfo {pages} {347} (\bibinfo
  {year} {1981})}\BibitemShut {NoStop}%
\bibitem [{\citenamefont {Linde}(1982)}]{linde}%
  \BibitemOpen
  \bibfield  {author} {\bibinfo {author} {\bibfnamefont {A.}~\bibnamefont
  {Linde}},\ }\href {\doibase http://dx.doi.org/10.1016/0370-2693(82)91219-9}
  {\bibfield  {journal} {\bibinfo  {journal} {Physics Letters B}\ }\textbf
  {\bibinfo {volume} {108}},\ \bibinfo {pages} {389 } (\bibinfo {year}
  {1982})}\BibitemShut {NoStop}%
\bibitem [{\citenamefont {Liddle}\ \emph {et~al.}(1994)\citenamefont {Liddle},
  \citenamefont {Parsons},\ and\ \citenamefont {Barrow}}]{liddle+}%
  \BibitemOpen
  \bibfield  {author} {\bibinfo {author} {\bibfnamefont {A.~R.}\ \bibnamefont
  {Liddle}}, \bibinfo {author} {\bibfnamefont {P.}~\bibnamefont {Parsons}}, \
  and\ \bibinfo {author} {\bibfnamefont {J.~D.}\ \bibnamefont {Barrow}},\
  }\href@noop {} {\bibfield  {journal} {\bibinfo  {journal} {Phys. Rev.}\
  }\textbf {\bibinfo {volume} {D50}},\ \bibinfo {pages} {7222} (\bibinfo {year}
  {1994})}\BibitemShut {NoStop}%
\bibitem [{\citenamefont {Dodelson}\ and\ \citenamefont
  {Hui}(2003)}]{dodelson}%
  \BibitemOpen
  \bibfield  {author} {\bibinfo {author} {\bibfnamefont {S.}~\bibnamefont
  {Dodelson}}\ and\ \bibinfo {author} {\bibfnamefont {L.}~\bibnamefont {Hui}},\
  }\href@noop {} {\bibfield  {journal} {\bibinfo  {journal} {Phys. Rev. Lett.}\
  }\textbf {\bibinfo {volume} {91}},\ \bibinfo {pages} {131301} (\bibinfo
  {year} {2003})}\BibitemShut {NoStop}%
\bibitem [{\citenamefont {Liddle}\ and\ \citenamefont {Leach}(2003)}]{LL}%
  \BibitemOpen
  \bibfield  {author} {\bibinfo {author} {\bibfnamefont {A.~R.}\ \bibnamefont
  {Liddle}}\ and\ \bibinfo {author} {\bibfnamefont {S.~M.}\ \bibnamefont
  {Leach}},\ }\href@noop {} {\bibfield  {journal} {\bibinfo  {journal} {Phys.
  Rev.}\ }\textbf {\bibinfo {volume} {D68}},\ \bibinfo {pages} {103503}
  (\bibinfo {year} {2003})}\BibitemShut {NoStop}%
\bibitem [{\citenamefont {del Campo}(2014)}]{delCampo}%
  \BibitemOpen
  \bibfield  {author} {\bibinfo {author} {\bibfnamefont {S.}~\bibnamefont {del
  Campo}},\ }\enquote {\bibinfo {title} {Springer handbook of spacetime},}\ \
  (\bibinfo  {publisher} {Springer Berlin Heidelberg},\ \bibinfo {address}
  {Berlin, Heidelberg},\ \bibinfo {year} {2014})\ Chap.\ \bibinfo {chapter}
  {Exact Approach to Inflationary Universe Models}, pp.\ \bibinfo {pages}
  {673--696}\BibitemShut {NoStop}%
\bibitem [{\citenamefont {Lidsey}\ \emph {et~al.}(1997)\citenamefont {Lidsey},
  \citenamefont {Liddle}, \citenamefont {Kolb}, \citenamefont {Copeland},
  \citenamefont {Barreiro},\ and\ \citenamefont {Abney}}]{lidseybig}%
  \BibitemOpen
  \bibfield  {author} {\bibinfo {author} {\bibfnamefont {J.~E.}\ \bibnamefont
  {Lidsey}}, \bibinfo {author} {\bibfnamefont {A.~R.}\ \bibnamefont {Liddle}},
  \bibinfo {author} {\bibfnamefont {E.~W.}\ \bibnamefont {Kolb}}, \bibinfo
  {author} {\bibfnamefont {E.~J.}\ \bibnamefont {Copeland}}, \bibinfo {author}
  {\bibfnamefont {T.}~\bibnamefont {Barreiro}}, \ and\ \bibinfo {author}
  {\bibfnamefont {M.}~\bibnamefont {Abney}},\ }\href@noop {} {\bibfield
  {journal} {\bibinfo  {journal} {Rev. Mod. Phys.}\ }\textbf {\bibinfo {volume}
  {69}},\ \bibinfo {pages} {373} (\bibinfo {year} {1997})}\BibitemShut
  {NoStop}%
\bibitem [{\citenamefont {Kinney}(2002)}]{kinney}%
  \BibitemOpen
  \bibfield  {author} {\bibinfo {author} {\bibfnamefont {W.~H.}\ \bibnamefont
  {Kinney}},\ }\href@noop {} {\bibfield  {journal} {\bibinfo  {journal} {Phys.
  Rev.}\ }\textbf {\bibinfo {volume} {D66}},\ \bibinfo {pages} {083508}
  (\bibinfo {year} {2002})}\BibitemShut {NoStop}%
\bibitem [{\citenamefont {Chongchitnan}\ and\ \citenamefont
  {Efstathiou}(2005)}]{me1}%
  \BibitemOpen
  \bibfield  {author} {\bibinfo {author} {\bibfnamefont {S.}~\bibnamefont
  {Chongchitnan}}\ and\ \bibinfo {author} {\bibfnamefont {G.}~\bibnamefont
  {Efstathiou}},\ }\href@noop {} {\bibfield  {journal} {\bibinfo  {journal}
  {Phys. Rev.}\ }\textbf {\bibinfo {volume} {D72}},\ \bibinfo {pages} {083520}
  (\bibinfo {year} {2005})}\BibitemShut {NoStop}%
\bibitem [{\citenamefont {Hoffman}\ and\ \citenamefont
  {Turner}(2001)}]{hoffman}%
  \BibitemOpen
  \bibfield  {author} {\bibinfo {author} {\bibfnamefont {M.~B.}\ \bibnamefont
  {Hoffman}}\ and\ \bibinfo {author} {\bibfnamefont {M.~S.}\ \bibnamefont
  {Turner}},\ }\href@noop {} {\bibfield  {journal} {\bibinfo  {journal} {Phys.
  Rev.}\ }\textbf {\bibinfo {volume} {D64}},\ \bibinfo {pages} {023506}
  (\bibinfo {year} {2001})}\BibitemShut {NoStop}%
\bibitem [{\citenamefont {Ramirez}\ and\ \citenamefont
  {Liddle}(2005)}]{ramirez}%
  \BibitemOpen
  \bibfield  {author} {\bibinfo {author} {\bibfnamefont {E.}~\bibnamefont
  {Ramirez}}\ and\ \bibinfo {author} {\bibfnamefont {A.~R.}\ \bibnamefont
  {Liddle}},\ }\href@noop {} {\bibfield  {journal} {\bibinfo  {journal} {Phys.
  Rev.}\ }\textbf {\bibinfo {volume} {D71}},\ \bibinfo {pages} {123510}
  (\bibinfo {year} {2005})}\BibitemShut {NoStop}%
\bibitem [{Note1()}]{Note1}%
  \BibitemOpen
  \bibinfo {note} {We will assume that $\phi (t)$ increases monotonically
  during inflation, and since $N$ and $t$ run in opposite directions, (\ref
  {solve-n}) implies that $H^\prime (\phi )<0$ during inflation.}\BibitemShut
  {Stop}%
\bibitem [{\citenamefont {{Planck Collaboration}}\ \emph
  {et~al.}(2015)\citenamefont {{Planck Collaboration}}, \citenamefont {{Ade}},
  \citenamefont {{Aghanim}}, \citenamefont {{Arnaud}}, \citenamefont
  {{Ashdown}}, \citenamefont {{Aumont}}, \citenamefont {{Baccigalupi}},
  \citenamefont {{Banday}}, \citenamefont {{Barreiro}}, \citenamefont
  {{Bartlett}},\ and\ \citenamefont {et~al.}}]{planck15}%
  \BibitemOpen
  \bibfield  {author} {\bibinfo {author} {\bibnamefont {{Planck
  Collaboration}}}, \bibinfo {author} {\bibfnamefont {P.~A.~R.}\ \bibnamefont
  {{Ade}}}, \bibinfo {author} {\bibfnamefont {N.}~\bibnamefont {{Aghanim}}},
  \bibinfo {author} {\bibfnamefont {M.}~\bibnamefont {{Arnaud}}}, \bibinfo
  {author} {\bibfnamefont {M.}~\bibnamefont {{Ashdown}}}, \bibinfo {author}
  {\bibfnamefont {J.}~\bibnamefont {{Aumont}}}, \bibinfo {author}
  {\bibfnamefont {C.}~\bibnamefont {{Baccigalupi}}}, \bibinfo {author}
  {\bibfnamefont {A.~J.}\ \bibnamefont {{Banday}}}, \bibinfo {author}
  {\bibfnamefont {R.~B.}\ \bibnamefont {{Barreiro}}}, \bibinfo {author}
  {\bibfnamefont {J.~G.}\ \bibnamefont {{Bartlett}}}, \ and\ \bibinfo {author}
  {\bibnamefont {et~al.}},\ }\href@noop {} {\bibfield  {journal} {\bibinfo
  {journal} {ArXiv e-prints}\ } (\bibinfo {year} {2015})},\ \Eprint
  {http://arxiv.org/abs/1502.01589} {arXiv:1502.01589} \BibitemShut {NoStop}%
\bibitem [{\citenamefont {{Coone}}\ \emph {et~al.}(2015)\citenamefont
  {{Coone}}, \citenamefont {{Roest}},\ and\ \citenamefont {{Vennin}}}]{coone}%
  \BibitemOpen
  \bibfield  {author} {\bibinfo {author} {\bibfnamefont {D.}~\bibnamefont
  {{Coone}}}, \bibinfo {author} {\bibfnamefont {D.}~\bibnamefont {{Roest}}}, \
  and\ \bibinfo {author} {\bibfnamefont {V.}~\bibnamefont {{Vennin}}},\ }\href
  {\doibase 10.1088/1475-7516/2015/11/010} {\bibfield  {journal} {\bibinfo
  {journal} {JCAP}\ }\textbf {\bibinfo {volume} {11}},\ \bibinfo {eid} {010}
  (\bibinfo {year} {2015})}\BibitemShut {NoStop}%
\bibitem [{\citenamefont {{Vennin}}(2014)}]{vennin}%
  \BibitemOpen
  \bibfield  {author} {\bibinfo {author} {\bibfnamefont {V.}~\bibnamefont
  {{Vennin}}},\ }\href {\doibase 10.1103/PhysRevD.89.083526} {\bibfield
  {journal} {\bibinfo  {journal} {Phys. Rev.}\ ,\ \bibinfo {eid} {083526}}
  (\bibinfo {year} {2014})}\BibitemShut {NoStop}%
\bibitem [{\citenamefont {Liddle}(2003)}]{Liddle}%
  \BibitemOpen
  \bibfield  {author} {\bibinfo {author} {\bibfnamefont {A.~R.}\ \bibnamefont
  {Liddle}},\ }\href@noop {} {\bibfield  {journal} {\bibinfo  {journal} {Phys.
  Rev.}\ }\textbf {\bibinfo {volume} {D68}},\ \bibinfo {pages} {103504}
  (\bibinfo {year} {2003})}\BibitemShut {NoStop}%
\bibitem [{Note2()}]{Note2}%
  \BibitemOpen
  \bibinfo {note} {The negative $\phi $ value is not problematic due to the
  even symmetry of $\protect \mathscr {H}$ and the $t\to -t$
  transformation.}\BibitemShut {Stop}%
\end{thebibliography}%


\end{document}